\documentstyle[epsfig]{mn}

\newif\ifAMStwofonts

\def\kms{\relax \ifmmode {\,\rm km\,s}^{-1}\else \,km\,s$^{-1}$\fi}
\def\ha{\relax \ifmmode {\rm H}\alpha\else H$\alpha$\fi}
\def\hb{\relax \ifmmode {\rm H}\beta\else H$\beta$\fi}
\def\hi{\relax \ifmmode {\rm H\,{\sc i}}\else H\,{\sc i}\fi}
\def\hii{\relax \ifmmode {\rm H\,{\sc ii}}\else H\,{\sc ii}\fi}
\def\h2{\relax \ifmmode {\rm H}_2\else H$_2$\fi}
\def\lha{\relax \ifmmode L_{{\rm H}\alpha}\else $L_{{\rm H}\alpha}$\fi}
\def\shi{\relax \ifmmode \sigma_{{\rm HI}}\else $\sigma_{\rm HI}$\fi}
\def\sh2{\relax \ifmmode \sigma_{{\rm H}_2}\else $\sigma_{{\rm H}_2}$\fi}
\def\degr{\hbox{$^\circ$}}
\def\arcmin{\hbox{$^\prime$}}
\def\arcsec{\hbox{$^{\prime\prime}$}}
\def\deg{\hbox{$^\circ$}}

\def\fdg{\hbox{$.\!\!^\circ$}}
\def\fs{\hbox{$.\!\!^{\rm s}$}}
\def\farcm{\hbox{$.\mkern-4mu^\prime$}}
\def\farcs{\hbox{$.\!\!^{\prime\prime}$}}
\def\degd#1.#2{ #1\fdg#2 }                 
\def\mind#1.#2{ #1\farcm#2 }               
\def\secd#1.#2{ #1\farcs#2 }               
\def\hhh{\ifmmode {\rm ^h}              
         \else {${\rm ^h}$}
         \fi}
\def\sss{\ifmmode {\rm ^s}              
         \else {${\rm ^s}$}
         \fi}
\def\hms#1h#2m#3s{                      
                  \relax
                  \ifmmode #1^{\rm h}\,#2^{\rm m}\,#3^{\rm s}
                  \else \hbox{$#1^{\rm h}\,#2^{\rm m}\,#3^{\rm s}$}
                  \fi
                 }
\def\dms#1d#2m#3s{                      
                  \relax
                  #1\degr\,#2\arcmin\,#3\arcsec 
                 }
\def\hmsd#1h#2m#3.#4s{                  
                      \relax
                      \ifmmode #1^{\rm h}\,#2^{\rm m}\,#3\fs#4
                      \else \hbox{$#1^{\rm h}\,#2^{\rm m}\,#3\fs#4$}
                      \fi
                     }
\def\dmsd#1d#2m#3.#4s{                  
                      \relax
                      #1\degr\,#2\arcmin\,#3\farcs#4
                     }
\def\mag{\relax                          
        \ifmmode ^{\rm m}
        \else $^{\rm m}$
        \fi
       }
\def\magd#1.#2{                          
              \relax
              \ifmmode #1^{\rm m}
                       \hskip-0.55em.\hskip0.22em#2
              \else \hbox{#1$^{\rm m}
                    \hskip-0.55em.\hskip0.22em$#2}
              \fi
             }

\ifoldfss
  \ifCUPmtlplainloaded \else
    \NewTextAlphabet{textbfit} {cmbxti10} {}
    \NewTextAlphabet{textbfss} {cmssbx10} {}
    \NewMathAlphabet{mathbfit} {cmbxti10} {} 
    \NewMathAlphabet{mathbfss} {cmssbx10} {} 
  \fi
  \ifAMStwofonts
    \ifCUPmtlplainloaded \else
      \NewSymbolFont{upmath} {eurm10}
      \NewSymbolFont{AMSa} {msam10}
      \NewMathSymbol{\upi}     {0}{upmath}{19}
      \NewMathSymbol{\umu}     {0}{upmath}{16}
      \NewMathSymbol{\upartial}{0}{upmath}{40}
      \NewMathSymbol{\leqslant}{3}{AMSa}{36}
      \NewMathSymbol{\geqslant}{3}{AMSa}{3E}

      \let\leq=\leqslant 
      \let\geq=\geqslant 
    \fi
  \fi
\fi 

\ifnfssone
  \newmathalphabet{\mathit}
  \addtoversion{normal}{\mathit}{cmr}{m}{it}
  \addtoversion{bold}{\mathit}{cmr}{bx}{it}
  \newmathalphabet{\mathbfit} 
  \addtoversion{normal}{\mathbfit}{cmr}{bx}{it}
  \addtoversion{bold}{\mathbfit}{cmr}{bx}{it}
  \newmathalphabet{\mathbfss} 
  \addtoversion{normal}{\mathbfss}{cmss}{bx}{n}
  \addtoversion{bold}{\mathbfss}{cmss}{bx}{n}
  \ifAMStwofonts
    \ifCUPmtlplainloaded \else
      \UseAMStwoboldmath
      \makeatletter
      \new@mathgroup\upmath@group
      \define@mathgroup\mv@normal\upmath@group{eur}{m}{n}
      \define@mathgroup\mv@bold\upmath@group{eur}{b}{n}
      \edef\UPM{\hexnumber\upmath@group}
      \new@mathgroup\amsa@group
      \define@mathgroup\mv@normal\amsa@group{msa}{m}{n}
      \define@mathgroup\mv@bold\amsa@group{msa}{m}{n}
      \edef\AMSa{\hexnumber\amsa@group}
      \makeatother
      \mathchardef\upi="0\UPM19
      \mathchardef\umu="0\UPM16
      \mathchardef\upartial="0\UPM40
      \mathchardef\leqslant="3\AMSa36
      \mathchardef\geqslant="3\AMSa3E

      \let\leq=\leqslant 
      \let\geq=\geqslant 
    \fi
  \fi
\fi 

\ifnfsstwo
  \newcommand{\etal}{\mbox{\em et al. }}

  \DeclareMathAlphabet{\mathbfit}{OT1}{cmr}{bx}{it}
  \SetMathAlphabet\mathbfit{bold}{OT1}{cmr}{bx}{it}
  \DeclareMathAlphabet{\mathbfss}{OT1}{cmss}{bx}{n}
  \SetMathAlphabet\mathbfss{bold}{OT1}{cmss}{bx}{n}
  \ifAMStwofonts
    \ifCUPmtlplainloaded \else
      \DeclareSymbolFont{UPM}{U}{eur}{m}{n}
      \SetSymbolFont{UPM}{bold}{U}{eur}{b}{n}
      \DeclareSymbolFont{AMSa}{U}{msa}{m}{n}
      \DeclareMathSymbol{\upi}{0}{UPM}{"19}
      \DeclareMathSymbol{\umu}{0}{UPM}{"16}
      \DeclareMathSymbol{\upartial}{0}{UPM}{"40}
      \DeclareMathSymbol{\leqslant}{3}{AMSa}{"36}
      \DeclareMathSymbol{\geqslant}{3}{AMSa}{"3E}

      \let\leq=\leqslant 
      \let\geq=\geqslant 
    \fi
  \fi
\fi 

\ifCUPmtlplainloaded \else
  \ifAMStwofonts \else 
    \def\upi{\pi}
    \def\umu{\mu}
    \def\upartial{\partial}
  \fi
\fi

\title[Constraints on the nuclear emission from the Circinus
galaxy: the torus]{Constraints on the nuclear emission of the Circinus galaxy: the torus}
\author[M. Ruiz \etal]{M. Ruiz$^1$\thanks{E--mail: mili@star.herts.ac.uk}, A. Efstathiou$^2$, D.M.
Alexander$^3$, J. Hough$^1$\\ 
$^1$Department of Physical Sciences, University of
Hertfordshire, Hatfield, Herts AL10 9AB, United Kingdom\\ 
$^2$Astrophysics Group, Imperial College, Blackett Laboratory, Prince Consort Road, London SW7 2BZ, United Kingdom\\
$^3$Penn State University, Astronomy and Astrophysics, Davey Laboratory, University Park, PA16802, USA}

\pagerange{\pageref{firstpage}--\pageref{lastpage}}
\pubyear{2001}
\begin{document}
\maketitle
\label{firstpage}
\begin{abstract}

In the context of the unified model of Seyfert galaxies, we 
use observations from the literature and a radiative transfer model
to investigate the near--IR to mm emission produced by the presumed torus in 
the Circinus
galaxy, from 2$\mu$m to 1.3mm. From the infrared SED modelling, we find that the 
total luminosity (L$_{IR}$) in this wavelength range,
consists of similar contributions
from the torus and starburst with a ratio of nuclear luminosity to 
starburst luminosity (L$_{NUC}$/L$_{SB}$) $\sim$ 0.8.

By using a similar torus model to that of NGC1068, 
{\it but without the conical dust}, 
we find an upper limit to the outer torus radius of $\sim$12pc with
a best fit of $\sim$2pc. The upper limit torus size estimated   
from the radiative transfer modelling is consistent with
the 16pc torus radius estimated from near--IR imaging polarimetry of Circinus.

\end{abstract}


\section{Introduction} 

Observational evidence for obscuring material in the centres of Active 
Galactic Nuclei (AGN) is now abundant. Obscuration is observed in 
the form of molecules
and dust, present in the centres of these galaxies on scales that vary from 
a fraction of a parsec to tens and even hundreds of parsecs (Malkan \etal 1998; 
Gallimore; Siebenmorgen \etal 1997; Greenhill \etal 1996). 
It is believed
that the obscuring material takes the form of a dusty ``torus'' which
blocks  and absorbs part of the radiation from the active nucleus. These
dusty tori play an essential role in unified theories of Seyfert galaxies 
and strongly support the unification models.

The unified model for Seyfert galaxies proposes that all types of Seyfert
galaxy are fundamentally the same, but, the presence of the dusty 
torus obscures the nucleus, including the broad line emission region 
in many systems. In this picture
the classification of Seyfert 1 or 2 depends on the inclination angle of the
torus to our line of sight (e.g. Antonucci and Miller 1985; Antonucci
1993). Strong support for this model has come from X--ray observations,
showing that Seyfert 2 galaxies are Compton thick, i.e. the nuclear radiation is 
absorbed by matter with column densities $>$10$^{24}$cm$^{-2}$ (Lawrence and Elvis 1982;
Maiolino \etal 1998a).

Direct imaging of the presumed torus is technically demanding due 
to its (predicted)
small size (e.g. Efstathiou, Hough and Young 1995, hereafter EHY95; Granato, Danese and
Franceschini 1997; Alexander \etal 1999). The most convincing
direct evidence for a Seyfert torus comes from observations of the HCN
molecule (Jackson \etal 1993), which traces dense molecular gas, and near--IR
imaging polarimetry (Young \etal 1996), both of NGC1068. These images show a
large nuclear structure of $\sim$200pc in extent, approximately
perpendicular to the [O\,{\sc iii}]5007 emission cone and radio lobe
emission. Indirect support for tori comes from ground--based narrow
band imaging (e.g. Wilson and Tsvetanov 1994), high resolution HST imaging
(e.g. Capetti \etal 1997; Falcke, Wilson and Simpson 1998) and
imaging--polarimetry (e.g. Ruiz \etal 2000; Lumsden \etal 1999; Tadhunter \etal
1999; Packham \etal 1997) which have shown a number of objects with
nuclear cone--like structures, presumably collimated by the torus.

The origin of the infrared continuum in a number of Seyfert 2 galaxies is 
clearly thermal
as indicated by silicate features such as the broad absorption at 10$\mu$m 
(Roche \etal 1991) and the 
various emission features from polycyclic aromatic hydrocarbons (PAHs), e.g. 
3.3, 6.2 and 11.3$\mu$m, which are normally associated with the presence of
hot stars in a starburst environment.
Dust grains, distributed in a toroidal geometry surrounding the 
central black hole and accretion disc, are believed to be responsible for the 
absorption of 
radiation from the central source and re--radiation of energy into longer wavelengths
giving rise to the observed
``IR bump'' longwards of 1$\mu$m and peaking
at mid--IR wavelengths (Rowan--Robinson and Crawford 1989). The
resultant radiation spectrum is dependent on the distribution of dust grain
temperatures ($\leq$ 2000K) and optical depth. Two additional components can add to the IR spectra of
active galaxies (Rowan--Robinson and Crawford 1989): galactic diffuse emission from 
large dust grains
heated by the interstellar radiation field in the galactic disc
 (cirrus, T$_{large-grains} \leq$ 40K ) and a starburst component
peaking at about 60$\mu$m. 

Thus, a radiative transfer model which takes into account 
dust grain features within a true toroidal configuration and starburst environment
is needed to correctly reproduce the near--IR to mm SED of the Circinus galaxy.

Here, we present a model for the IR continuum emission of the
Circinus galaxy which successfully fits the observations from 2$\mu$m to 
mm wavelengths. 
We constrain the size and flux of the dusty
AGN torus and estimate the contribution of the starburst emission to total 
IR emission. 
Section 2
presents the data compilation and sections 3 and 4 present the model and 
the fitting procedure. Section 5 discusses the model and its 
implications and conclusions are summarised in section 6.

\section{The Data}

\begin{table}
\begin{minipage}[t]{3.0in}
  \caption[]{IR--mm data points\footnote{Collected from the literature. Data for the 7--13$\mu$m range are shown in Fig 1}}
  \label{tab:table}
    \leavevmode   
    \footnotesize
    \begin{tabular}[h]{cccc}
      \hline \\[-5pt]
$\lambda$ & flux & beamsize  &reference\footnote{References:
(1) Maiolino \etal (1998b); (2) Sturm \etal (2000); 
(3) Moorwood and Glass (1984);
(4) IRAS, Moshir \etal (1992); (5) Ghosh \etal (1992); 
(6) Siebenmorgen \etal (1997) }\\
 ($\mu$m )       &  (Jy)& & \\
           &       &            &     \\
  2.2 \footnote{nuclear measurement}     & 0.022 &  0.15 arcsec         &  1\\
  3.0       &    1.1                             &  14$\times$20 arcsec$^2$& 2	\\
  3.28\footnote{PAH feature peak}      & 2.0   & 14$\times$20 arcsec$^2$    &  2  \\
  3.8$^b$   & 0.777 &  0.3 arcsec         &  6   \\
  4.8$^b$     & 1.832 &  0.3 arcsec         &  6   \\
  6.2$^c$    &   8.5 &  14$\times$20 arcsec$^2$    &  2  \\
  7.7$^c$     &  20.0     &   14$\times$20 arcsec$^2$   &  2   \\
  8.6$^c$       & 10.0  &  14$\times$20 arcsec$^2$    &  2   \\
  10.3$^b$     &  6.32&   1.5 arcsec   &  6  \\  
  11.3$^c$     & 20.0  & 14$\times$20 arcsec$^2$     &  2    \\
  12       &   19  &  90 arcsec &  4    \\
  20        &   45  &  14$\times$20 arcsec$^2$ & 2   \\
30        &   100  &  14$\times$20 arcsec$^2$ & 2   \\
40        &   200  &  14$\times$20 arcsec$^2$ & 2    \\
  25       &    65 &  90 arcsec &  4   \\
  60       &   280 &  90 arcsec &  4  \\
 100       &   340 &  90 arcsec &  4    \\
 150\footnote{5$\sigma$ upper limit}      &    $<$150 & 40 arcsec  &  5    \\
 1300     &  0.248 &  23 arcsec &  6 \\

\hline \\[-5pt]

      \end{tabular}
\end{minipage}
\end{table}

The continuum and PAHs data points have been compiled from a variety of sources in 
the literature and are presented in Table 1. The data sources will be 
described below. 
There is a large range of sizes in the observational apertures,
from arcsec to arcminute beamsizes, equivalent to physical sizes of  $<$5pc
and $>$200pc (for Circinus, 1 arcsec corresponds to 20pc); whenever possible,
data from the smallest available aperture were used
to minimise the contribution from galactic disk emission.

In the near--IR, below 2$\mu$m, the stellar contribution is quite significant,
and
to study the nuclear emission, the starlight 
contribution has to be subtracted from the integrated flux. The data
point at 2.2$\mu$m shown in Table 1 corresponds to the  deconvolved
non--stellar measurement as presented by Maiolino \etal (1998b) and derived from a 
0.15arcsec aperture. This measure will set a limit on the size of 
the nuclear non--stellar source (see section 2.1). 

In the mid--IR, between 3 and 20$\mu$m, the continuum data points at 3.8 and 4.8$\mu$m,
are provided by Siebenmorgen \etal (1997) who carried out speckle interferometry in 
the L'(3.8$\mu$m) and
M(4.8$\mu$m) bands with a 0.3arcsec aperture. The speckle observations are 
insensitive to larger scale components, e.g. starlight, and predict more directly 
the nuclear non--stellar contribution (see section 4). 
Other continuum data points over this wavelength range are from the ISO--SWS 
spectrum presented in Sturm \etal (2000).

Line information for the PAH features at 3.28, 6.2, 7.7, 8.6 and 11.3$\mu$m 
were directly measured from the ISO--SWS spectrum in Sturm \etal (2000). 
The reported values in Table 1 correspond
to the peak of these emission lines. ISO apertures are large enough 
($\sim$24arcsec$^2$) to contain much
of the circumnuclear star forming regions as well as the nuclear emission. 
The ISO spectrum shows that the region
between 5 and 12$\mu$m is dominated by PAH features, characteristic of star
formation activity (see Fig 1), and it is remarkably similar to
those of pure starburst galaxies also observed by ISOPHOT (Siebenmorgen, Kr\"ugel and Zota 1999). 

The spectrum corresponding to the Si absorption feature centered at 9.7$\mu$m 
is shown in more detail in Fig 1. It was provided
by P. Roche and published by Roche \etal (1991). This spectrum was taken with the UCL
spectrometer and used an aperture of 4.3arcsec. For comparison also shown in Fig 1 is the 
mid-IR ISO--SWS spectrum of Circinus taken with the much larger aperture of 
14$\times$20arcsec$^2$ (Sturm \etal 2000). 

The N(10.3$\mu$m) band continuum data point, is also provided by 
Siebenmorgen \etal (1997) and
measured through a 1.3arcsec slit. It is interesting to note that a larger 
aperture measurement at 10.3$\mu$m by Moorwood and Glass (1984) 
with a 5arcsec aperture is 
very similar to the smaller aperture data of Siebenmorgen \etal (1997). 
This is a good indication of the 
compact nature of the central source at this wavelength. 
The continuum data points at 20, 30 and 40$\mu$m were taken from the ISO--SWS 
spectrum (Sturm \etal 2000).
The ISO SWS spectrum at 25$\mu$m agrees 
within 20$\%$ of the {\it{IRAS}} flux.

The far--IR continuum observations were collected from the {\it{IRAS}}~ 
large aperture observations (90arcsec) at 12, 25, 60 
and 100$\mu$m as reported by Moshir \etal (1992). 
The continuum data point at 150$\mu$m was taken from the analysis by Ghosh \etal (1992)
which corresponds to an aperture of 40arcsec. This measurement is only a 5$\sigma$
 upper limit as
Circinus was not detected at this wavelength.

At millimeter wavelengths, Circinus has been detected at 1.3mm with an aperture 
of 23arcsec as reported by 
Siebenmorgen \etal (1997). Maps of the millimeter continuum emission show Circinus as an
unresolved source.

Clearly, the large aperture observations  
are contaminated with light from various emission sources 
particularly from 
the extended galactic disc; thus, special care is needed when fitting the 
large aperture data points to the individual emission components in the radiative
 transfer model (see next section).
\begin{figure*}
\begin{center}
\centerline{\epsfig{file=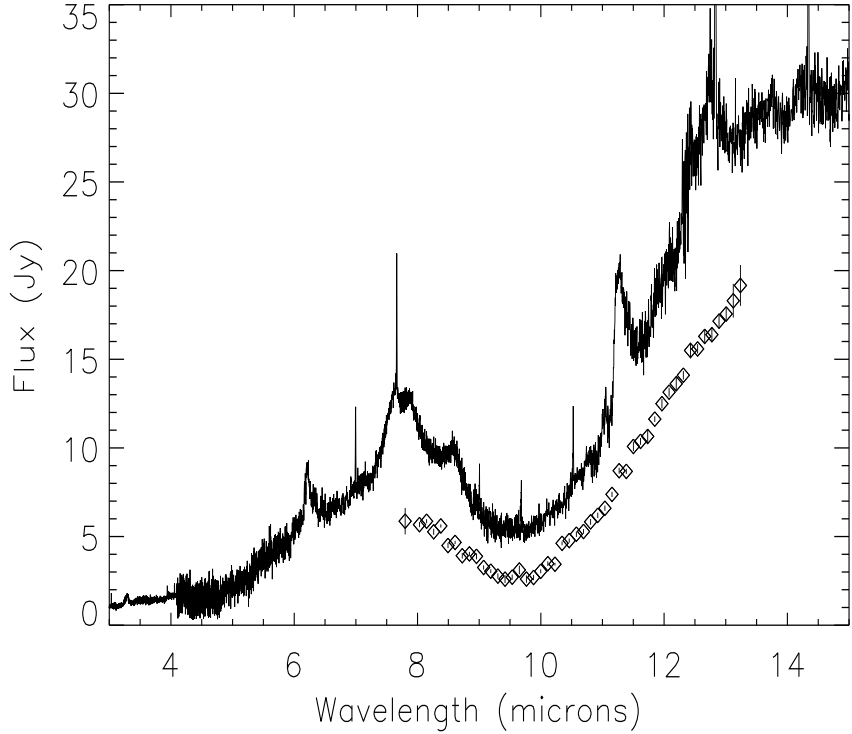,width=13cm,height=11cm,angle=0}}

\caption{
Circinus Mid IR spectrum. Solid line corresponds to the Circinus 
ISO-SWS (Sturm \etal 2000)
spectrum and square open points correspond to the Roche etal data. The ISO spectrum
is dominated by emission features from PAHs as the ISO-SWS spectrum was
taken with a large aperture (14$\times$20arcsec$^2$ ). Note the lack of these emission features in the Roche etal data (aperture $\sim$ 4.2arcsec).}

\end{center}
\end{figure*}

\subsection{Constraints from Observational Data}

 In the K band (2.2$\mu$m), the source is
clearly unresolved in a 0.3arcsec aperture (6pc on source).
By deconvolving the nuclear K--band surface brightness radial distribution 
with a stellar profile and a nuclear PSF, 
Maiolino \etal (1998b) set a constraint on the size of the nuclear non--stellar
source of $<$1.5pc.
Similarly, at 3.8 and 4.8$\mu$m the source is also observed to be 
unresolved at 0.7arcsec
resolution and is possibly $<$0.3arcsec in size, putting upper limits of 
6pc (Siebenmorgen \etal 1997).
Clearly, at wavelengths $\leq$5$\mu$m most of the emission comes from a
central 
source with a size $\leq$6pc. This size scale almost
certainly corresponds
to the inner radius of the proposed dusty torus, as the near--IR emission 
is most likely to result from
hot dust grains (T$\leq$1300K) heated by the intense radiation from the non--stellar
central source. {\it{Thus, in our model, the near--IR data points at $<$5$\mu$m are 
assumed to be totally dominated by the emission from the torus 
component.}}

Observations at 10$\mu$m reveal that the central source size is possibly resolved with 
a size of 26pc
whilst at 20$\mu$m it is unresolved and $<$30pc in size (Siebenmorgen
\etal 1997). 
As mentioned in the previous section, the emission detected with the larger aperture of
Moorwood and Glass (1984) at 10$\mu$m does not seem very different from that of smaller aperture by Siebenmorgen \etal (1997).
At wavelengths larger than  20$\mu$m, the IR emission most likely results from 
a more extended region,  and there is possibly
a significant contribution from the inner circumnuclear starburst ($\sim$ 40pc in radius).

The spatial resolution analysis at 60, 100 and 150$\mu$m by Ghosh \etal (1992) reveals
a deconvolved size of Circinus of 40arcsec at 50$\mu$m. However, at these longer wavelengths, 
the resolution 
is too poor to put useful
constraints on the size of the emitting source. 

The spectrum corresponding to the Si feature at 9.7$\mu$m (Fig 1)
was taken with an on--source
equivalent aperture of $\sim$80pc (Roche \etal 1991). 
The detection of a number of CO stellar absorption features in 
a 4.4$\times$6.6arcsec$^2$ aperture gives strong evidence for circumnuclear
star--formation on a region few tens of parsecs from the nucleus 
(Oliva \etal 1994,
Maiolino \etal 1998b), thus, the Si 
feature cannot be solely modelled by dust self--absorption 
from the torus component, but, a starburst
contribution will have to be included when fitting this feature with 
the SED model. 

The long wavelength data points and PAH features
(IRAS and ISO data), which correspond to large aperture observations 
(see Table 1), are hence dominated 
by the larger scale starburst emission observed in the 
circumnuclear ring
at $\sim$200pc from the nucleus. Thus, the data points corresponding to
IRAS and ISO data are fitted with a starburst component.  

\section{The General Infrared model}
The model used is the radiative transfer model of Efstathiou \etal as
described  in detail in Efstathiou and Rowan--Robinson (1995, hereafter ER95),
EHY95, and Efstathiou, Rowan--Robinson 
and Siebenmorgen (2000, hereafter ERS). This radiative transfer model
has been successful in fitting the IR
continuum of NGC1068 (EHY95) and Centaurus A
(Alexander \etal 1999).
The proposed model for the SED of the Circinus galaxy is a combination of the
following two components described below.

\subsection{Torus}
As concluded in ER95, the most likely geometry for the obscuring material in the center of AGNs, 
is that of a ``tapered disc''. Other proposed geometries for the torus 
are ``flared discs'' (Granato and Danese 1994; Efstathiou and Rowan--Robinson 1990 (hereafter ER90)) or cylinders (Pier and Krolik 1992; 1993; Taniguchi and Murayama 1998).
However, the shape of the IR 
continuum spectrum of type 1 and type 2 objects, the appearance of the 10$\mu$m
silicate features and the statistics of the two types of Seyfert galaxies,
give strong support for a tapered disc geometry 
(see ER95 for detailed discussion).
Indeed, there is further confirmation of the validity of the
tapered disc geometry to realistically represent the nuclear emission of tori in AGNs. This comes from
the fitting of IR data to a sample of Seyfert galaxies with the EHY95 models of Figure 3. They have been successfully used to 
fit nuclear (non--stellar) 
near--IR observations for a sample of Sy1s and Sy2s (Alonso--Herrero \etal 2001) showing 
that the near--IR nuclear emission from Seyfert galaxies can be  fitted with the tapered disc geometry tori of EHY95.
We therefore assume 
this geometry for the Circinus torus.  
As demonstrated in ER90, the orientation of the plane of the system
to the line--of--sight is a very important parameter when fitting 
the observed spectra, since only a small difference in this
inclination parameter, $\theta_V$,
produces a significant variation in the emergent spectrum. 
Another important parameter that determines the shape of
the IR continuum, is the half--opening angle of the torus, $\Theta$. 
In general, these parameters are not known for most objects, 
but fortunately, we 
have  a good idea of their values from spectropolarimetry 
(Alexander \etal 2000) and IR--imaging 
polarimetry of
Circinus (Ruiz \etal 2000) which gives  
$\theta_V$$\sim$40 degrees (measured from the axis of the torus to the line of sight)
and $\Theta$$\sim$45 degrees. Although the emission line cones 
observed in Circinus and many other AGNs
offer a measure of the opening angle of the torus, it is common to find that the 
conical emission ``opens--up'' as light travels and scatters in the circumnuclear medium. Thus, 
we should keep in mind that $\Theta$ can be smaller than that derived 
from emission line cones 
(see section 4).

Other parameters that
will affect the shape of the emergent SED spectrum are:
the ratio of the inner and outer radii ($r_1/r_2$), the ratio of the 
height to the outer radius ($h/r_2$), the equatorial optical depth
($^{eq}$A$_{UV}$), the dust sublimation temperature ($T_1$) and the 
radial dependence of the density distribution $r^{-\beta}$.
To take into account the effects of dust in the nuclear regions
of Circinus, we assume that there is an additional extinction (i.e. in 
addition to that produced by the torus). 
Marconi \etal (1994) found A$_V\geq$ 20mag 
towards the nucleus.  This limit follows from
the fact that we can see the nucleus at 1.25$\mu$m but are unable to do 
so at 7000\AA, thus, they only set a lower limit for the nuclear extinction.
However, we find that A$_V$=25~mag is required to successfully fit the near--IR 
data. This value
is also close to that found by Moorwood and Glass (1984) from the depth of
silicate feature at 9.7$\mu$m and 
 is also consistent with that estimated by Maiolino \etal (1998b). 
Based on H and K band observations, they set a minimum 
extinction A$_V$ $>$~12mag toward the nucleus of Circinus. 
From their images, it is
clear that this obscuration is caused by the large amounts of dust as
seen in their 
H--K colour map and it is due to the presence of a nuclear gas bar.
Their estimated A$_V$ however is only a lower limit and we expect a larger 
value since the extinction is most 
likely to be due to an inhomogeneous distribution of dust, gas and stars.
The parameters assumed for the torus model are listed in Table 2.
\begin{table}
\begin{minipage}[t]{5.5in}
  \caption{Model parameters}
  \label{tab:table}
    \leavevmode   
    \footnotesize
    \begin{tabular}[h]{lc}
\hline \\[-5pt]
\hspace{1.2in}
\vspace{0.1in}
        Best fit torus parameters\\
      system inclination\footnote{measured from the plane of the torus to the line of sight}, $\theta_V$, ($\deg$)          & 50 \\
      cone opening half-angle, $\Theta$, ($\deg$)   &  30\\
      $^{eq}$A$_{UV}$ (mag)          	&   1000\\
      $^{eq}$A$_{V}$ (mag)           & 200          \\
      $^{los}$A$_{V}$ (mag)          &    83\footnote{$^{eq}$A$_{V}$ corrected for the inclination of the torus to the line of sight}             \\
      $T_1$(K)\footnote{dust sublimation temperature at torus inner radius} &    1000 \\
      $\beta$\footnote{value determined as discussed in EHY95}                       	&   1      \\
      additional A$_V$\footnote{due to circumnuclear bar}(mag)               	&   25    \\
                       &          \\
\hline \\[-5pt]
\hspace{1.0in}
\vspace{0.1in}
      Best fit starburst parameters\\
	
      Starburst e-folding time              	&   	10 Myr\\
	Starburst age                        		&   	26 Myr\\
	$\tau_V$, initial GMC optical depth     &   	50\\
      A$_V$ in ISM (mag)           			& 	5\\
      A$_V$ through Galactic disk (mag)           	&	 1.5\\
     \hline
      \end{tabular}
\end{minipage}
\end{table}

\subsection{Starburst}
The radiative transfer models of starburst galaxies are described 
in detail in ERS. The models consider
a starburst galaxy as an ensemble of optically thick giant molecular
clouds centrally illuminated by recently formed stars. 
The stellar population is determined according to the Bruzual and Charlot (1993)
stellar population synthesis models. The grain model used is that of 
Siebenmorgen and Kr\"ugel (1992) and the emission of transiently heated 
dust grains is calculated according to their method. In the models, the effects of multiple scattering is taken into account. The models relate the observed properties of a galaxy to its age and star formation history.
These models have 
successfully matched the observational characteristics of M82 and other starburst 
galaxies (ERS).
Various populations of dust particles are considered:\\
(i) Mathis, Rumpl and Nordsieck (1977). These are large particles with a 
size distribution of classical grains\\
(ii) small graphite grains, to account for the extinction bump at around 
2175\AA ~(Draine ~1989)\\
(iii) PAH molecules, to explain the near--IR and mid--IR emission bands\\
To take into account the effects of dust in the starburst regions 
of Circinus, we extinct the starburst emission by A$_V$=5mag, as measured from  narrow  
emission line ratios (Oliva \etal 1994).
This is lower than the additional extinction assumed to the nucleus (section 3.1), as the starburst 
is expected to be outside the circumnuclear molecular bar.

\section{Model fitting}
\begin{figure*}
\begin{center}
\centerline{\epsfig{file=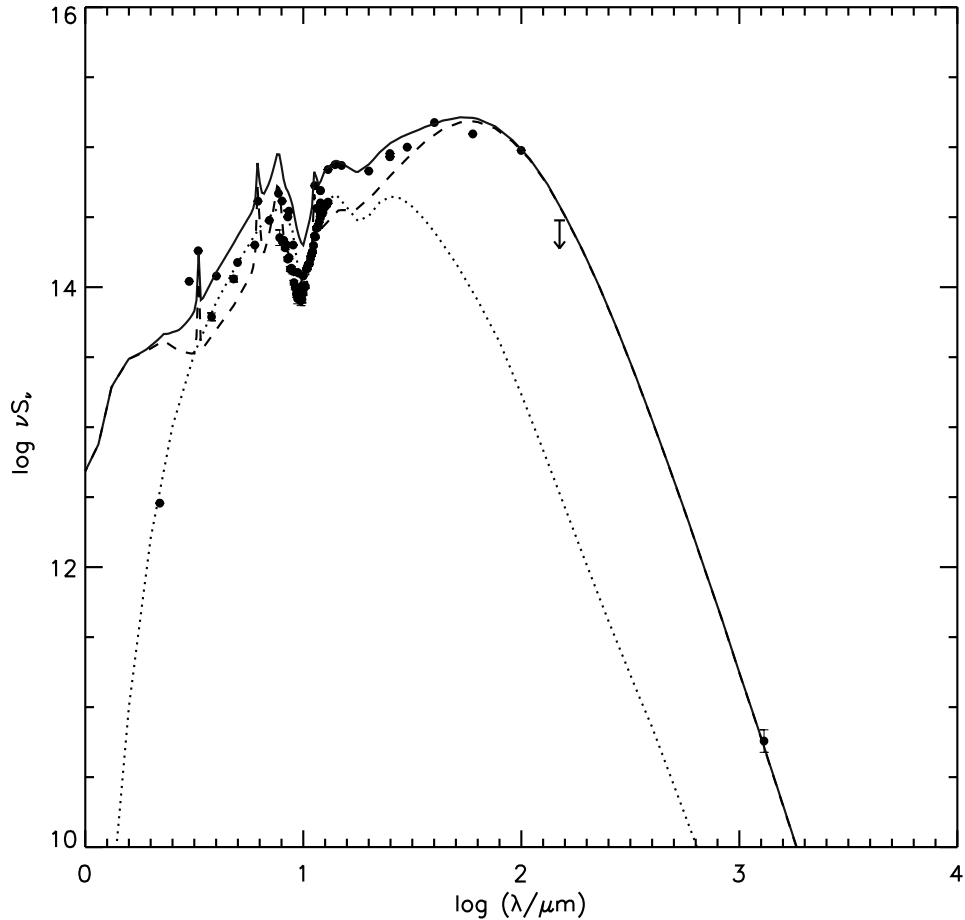,width=13cm,height=14cm,angle=90}}
\caption{IR SED for the Circinus galaxy. The SED is fitted with a starburst and a torus model. 
Dark circles are the data points from Table 1 and Figure 1. Units for $\nu S_{\nu}$
are 10$^{-26}$W/m$^2$. Dotted line corresponds to the torus 
component, dashed line is the starburst component and the solid line represents the total fit. 
The model parameters assumed for the torus and starburst are given in Table 2.}

\end{center}
\end{figure*}
As described in section 3, the data points will be fitted to the SED model
components according to the observational constraints described in section 2.1. 
The critical 
wavelength range is likely to be where the torus and starburst components are equally 
important, that is
in the 10--30$\mu$m region. Here, the strong absorption feature is indicative of the 
presence of dust in the form of silicates which are likely to be present in
a dusty environment such as the
torus, however, the observational beam for this data is large enough to 
cover a  
region which also includes star formation. Thus, contributions from both 
the torus and starburst components are important in this wavelength range.
On the other hand, given that the near--IR points are to be modelled by the torus 
alone (see section 2.1), this sets a strong constraint on the number 
of possible models to be 
tested for the torus.  Similarly, in the long wavelength range and 
large aperture data points which include the PAHs features, 
the starburst component dominates, 
and only a number of starburst models can be applied. Thus, 
the overall fitting procedure is much better constrained and the number of free 
parameters reduced once the short and long wavelength ranges are considered.

The torus input parameters that reproduce the near--IR points are 
listed in Table 2. The best fits were obtained with a cone 
half--opening angle of 30 degrees,
the same as the best fit obtained for the NGC1068 torus model (EHY95).
One important difference with the NGC1068 model is that there is no 
requirement to include a contribution from hot dust in the torus cone.
Indeed, if this is included, then it is very difficult to obtain a fit to
the near--IR data points.
The starburst parameters which reproduce the PAHs features and large scale
emission successfully are also listed in Table 2. Table 3 presents the 
derived geometrical 
characteristics for the best fit torus in Circinus. 
In Figure 2 we show the best model for the SED of the Circinus galaxy
which is the addition of the torus and starburst models. Note that the solid line
(representing the total emission from the galaxy) exceeds the mid--infrared data of
Roche \etal (1991). This is not surprising as the starburst which 
contributes about half of the emission is more extended that the 4arcsec 
aperture of Roche \etal (1991). Thus, the final SED for the 10$\mu$m 
absorption feature should not fit the 9.7$\mu$m data of Roche \etal 1991 (Fig 1).

Unfortunately, there is a serious lack of small aperture (nuclear)
data in the range of 30--60$\mu$m. As a consequence, we are unable to rule out 
 tori up to $\sim$12pc although increasing the
torus outer radius further would produce far too much 30--60$\mu$m flux.
Future observations in this critical wavelength range, will put extra constraints on the torus 
model, and the torus size would be determined more accurately. 

It is interesting to note in Figure 2 the overall larger contribution to the total IR emission of
the starburst  compared to the torus component. This is indicated by the 
ratio of the nuclear luminosity --torus component-- to starburst 
luminosity --starburst component-- ($L_{NUC}$/$L_{SB}$) $\sim$ 0.8 (see next section).
\begin{table}
\begin{minipage}[t]{5.5in}
  \caption{Torus characteristics}
  \label{tab:table}
    \leavevmode   
    \footnotesize
    \begin{tabular}[h]{lc}
\hline \\[-5pt]

$r_1/r_2$         & 0.05 \\
$h/r_2$           & 0.5  \\
inner radius ($r_1$, pc)  &0.12  	\\
outer radius ($r_2$, pc)  &2.4 	\\
height ($h$, pc)        &1.2 	\\
                   &          \\
\hline
      \end{tabular}
\end{minipage}
\end{table}
 
\section{Discussion}

Our SED model shows that,
overall, the starburst contribution to the total IR luminosity, $L_{IR}$
slightly dominates over the torus contribution. A direct result of this, 
is the relatively small
size of the torus in Circinus, with a best fit of $\sim$2pc for
 the outer radius, 
comparable to the torus outer radius of 1.8pc modelled for Cen A
(Alexander \etal 1999) but significantly smaller
than the $\sim$90pc torus for NGC1068 (Young \etal 1996 and
EHY95). This result is smaller than that calculated from 
the near--IR imaging polarimetry
modelling of Ruiz \etal (2000) which suggests that the torus is 
approximately 16pc in radius, although consistent with our upper limit of 
 $\sim$12pc. 

So far, by using the radiative transfer models of Efstathiou \etal we have 
been able to model the size of the torus outer radius for three objects: 
NGC1068 (EHY95), CenA (Alexander \etal 1999) 
and Circinus (this work). Although the statistics are small, it 
is interesting to note that all three AGN can be modelled with tori of very similar parameters. 
This suggests that 
whatever the mechanism that forms these tori in an AGN--SB environment, it
is such that the size of the tori scales with the luminosity of the central 
source but the rest of the torus parameters (Table 2) are scale invariant. 
For example, the hard X--ray continuum, which is considered to be 
a good indicator of the central source strength, indicates that the
Circinus galaxy is a relatively low luminosity AGN (Matt \etal 1996), while NGC1068
is $\sim$ 100 times more powerful (Turner \etal 1997), with a torus radius $\sim$100pc. The predicted torus
radii
scales as $L^{1/2}$, reflecting the fact that both objects can be
fitted with very similar radiative transfer models.

Of great debate regarding the nature of the Circinus galaxy is the ratio of the 
AGN to starburst emission ($L_{NUC}$/$L_{SB}$), that is, the relative 
contribution of the 
torus and starburst components to the total $L_{IR}$. 
Previous predictions from IR modelling differ widely. As already 
mentioned, Rowan--Robinson and Crawford (1989) found a 20$\%$ contribution of 
the nuclear component to the total $L_{IR(10-100\mu m)}$ and $L_{NUC}$/$L_{SB}$ $\sim$0.3 while this work
suggests a $L_{NUC}$/$L_{SB}$ of 0.8. On the other hand, from the Br$\alpha$ 
luminosity, Moorwood \etal (1996) estimate that this ratio is
perhaps as high as 9.
Our result is in very good agreement
with that suggested by
Maiolino \etal (1998b) who estimate from K band observations a $L_{NUC}$/$L_{SB}$ ratio of 0.9 
for the region $<$200pc. Similarly, we agree with the ratio derived
from the inferred ionising
continuum required to produce the highly ionised IR lines suggesting a ratio
closer to 1 (Moorwood \etal 1996).
We believe that the differences are caused by a number 
of pitfalls.
For example,  Rowan--Robinson and Crawford (1989) modelled the far--IR (10-100$\mu$m) {\it{IRAS}}
data with a three component model, a starburst, a Seyfert and a disc cirrus--like 
component. 
This model assumed a spherically symmetric dust distribution with a power--law 
density distribution, $n(r) \sim r^{-\beta}$. In this model, radiation from the 
polar regions of the torus is absorbed by the spherical dust cloud.
Thus, this model will naturally
under--estimate the nuclear component, as their low $L_{NUC}$/$L_{SB}$ ratio shows. However, there is now ample observational 
evidence that favours the existence of axially symmetric distributions of dust 
(Efstathiou and Rowan--Robinson 1990, ER95). 
On the other hand the large $L_{NUC}$/$L_{SB}$ ratio derived  by  Moorwood \etal 
(1996) suggests a strongly dominant AGN over the starburst. This 
was derived from the comparison of near--IR recombination lines in Circinus and M82; from a Br$\gamma$ map (Moorwood and Oliva 1994), they assume that 50$\%$ of 
the recombination lines are produced in the nucleus
of Circinus. However, this is likely to be an over--estimation of the strength of  Br$\gamma$
in the nucleus since,,more recently, a more accurate measurement of the
Circinus nuclear recombination lines (Maiolino \etal 1998b) has shown that 
only $\sim$10$\%$ of the 
total recombination lines flux is produced in the nucleus of Circinus. 
Thus, assuming the 
standard relation between Br$\alpha$ flux and total IR luminosity in starburst 
galaxies, we can
determine the contribution of the starburst to $L_{IR}$. We find $L_{SB}$ $\sim$ 
8.5$\times10^9L_{\odot}$, so $L_{NUC}$/$L_{SB}$ $\sim$ 1.2 for an AGN luminosity of $10^{10}L_{\odot}$.
 This is in good agreement with the ratio we
find from the SED modelling.

Radio observations suggest
that at these wavelengths, the starburst emission is the dominant component (Forbes and Norris, 1998) and Circinus
would certainly be classified as a starburst galaxy based purely on radio
emission. Indeed, it is interesting to note that the FIR--radio 
correlation established by Helou, Soifer and Rowan--Robinson (1985) which is 
followed by normal spiral and starburst galaxies, is closely followed by the Circinus 
galaxy, confirming Forbes and Morris result. 
However, it is almost certain that this correlation is the result of large scale 
star--formation activity unrelated to the Seyfert activity 
which generates both the synchrotron radio emission and the thermal FIR emission.
Given that the Circinus galaxy
has strong star--formation activity on small and large scales, 
it is not surprising that this galaxy follows the FIR--radio correlation.
Radio observations on arcsec resolution can not easily distinguish between
Seyfert and starburst activity. The radio emission from both starburst regions
and Seyfert nuclei have similar spectra and morphology, and the steep spectrum of even
the compact cores seen in VLA maps of Seyferts suggests that they might contain a
significant nuclear starburst component (Norris \etal 1992). On the other hand,
long--baseline radio interferometry is sensitive only to compact, high brightness  
objects and is an ideal tool to discriminate over the two spatial scales.
Unfortunately no radio interferometry is available for the Circinus galaxy and we 
can not clearly determine the nature of the central source from present
radio data.

Nevertheless, there is a clear large variation in the 
predicted $L_{NUC}$/$L_{SB}$ ratios for Circinus,
from starburst-dominated at FIR wavelengths, to AGN-dominated in the near and mid IR.
Indeed,
NGC1068, the prototypical Seyfert 2 galaxy shows a similar behaviour. 
The mid--IR(5-16$\mu$m) spectrum of NGC1068 is 
$\sim$85-95$\%$ dominated by the AGN, and $L_{NUC}$/$L_{SB}$ $\sim$ 3. 
In the FIR however, where extended emission dominates, the nucleus does
 not contribute more than 25$\%$ to the total IR flux at 450$\mu$m 
(Le Floc'h \etal 2001).
Clearly, determination of $L_{NUC}$/$L_{SB}$ is strongly dependant on the spectral region where it is calculated.

Our SED modelling has the advantages over other predictions of 
having a more complete model for the SED, including 
the emission of PAH features and a more realistic torus model. Additionally we
have used a large set of recently published data, from NIR to mm wavelengths.
More importantly, we pay special attention to the fit of data points 
from the various aperture sizes which give important constraints to the size of the
emitting source. 
The undoubted composite nature of the Circinus galaxy is shown in our IR SED model which
predicts a galaxy dominated by the AGN torus at near and mid--IR wavelengths 
while at far--IR to mm wavelengths 
the starburst activity is the dominant one.

\section{Conclusions}

We can summarise our results as follows:

\begin{itemize}

\item The case of the Circinus galaxy provides strong support for the dusty torus model of active galaxies 
which further support the unification scheme, that is, 
thick dusty tori with a tapered disc geometry surrounding the central engine.

\item Our best fit for the outer radius of the torus in Circinus is $\sim$2pc, but a radius as large as 12pc can not be ruled out.

\item We confirm the composite nature of the Circinus galaxy, which is dominated  
by the
AGN torus emission in the near--IR but largely dominated by starburst phenomena in the far--IR and mm
wavelengths. The $L_{IR}$ is slightly dominated by the starburst emission and the ratio
$L_{NUC}$/$L_{SB}$ $\sim$ 0.8.

\end{itemize}

\section{Acknowledgements} 
We thank P.Roche for providing the mid--IR spectrum of Circinus and D.Sturm for providing the ISO SWS spectrum. We thank 
A. Alonso--Herrero and the 
anonymous referee for useful comments.
MR and AE thank PPARC for support through a postdoctoral assistantship. DMA
thanks the TMR network (FMRX-CT96-0068) for support through a postdoctoral
grant while working on this project.  This research has made use of 
the NASA/IPAC Extragalactic Database
(NED), which is operated by the Jet Propulsion Laboratory, California
Institute of Technology, under contract with NASA.

\bsp

\label{lastpage}

\end{document}